\newcommand{\rev}[1]{{{#1}}}

\documentclass[journal
]{IEEEtran}

\usepackage{amsmath,amssymb,epsfig,epstopdf,siunitx,pgfplots}

\pgfplotsset{compat=1.9}

\begin{document}
	\newlength\figureheight
	\newlength\figurewidth

\title{Uncertainty Quantification of Oscillation Suppression during DBS in a Coupled Finite Element and Network Model}

\author{Christian~Schmidt$^1$,~
        Eleanor~Dunn$^2$,~
        Madeleine~Lowery$^2$,~\IEEEmembership{Member,~IEEE}
				and~Ursula~van~Rienen$^3$,~\IEEEmembership{Member,~IEEE}
\thanks{The work was supported by the DFG (German Science Foundation) Research Training Group 1505 "welisa" and ERC Consolidator grant (646923).}
\thanks{\textit{Asterisk indicates corresponding author.}}%
\thanks{$^1$$^\ast$C. Schmidt and U. van Rienen are with the Institute of General Electrical Engineering, University of Rostock, 18059 Rostock, Germany
        {\tt\small cschmidt18057 at gmail.com}}%
\thanks{$^2$E. Dunn and M. Lowery are with the School of Electrical, Electronic and Communications Engineering, University College Dublin, Dublin 4, Ireland
        }%
}

\markboth{Journal of \LaTeX\ Class Files,~Vol.~13, No.~9, September~2014}%
{Shell \MakeLowercase{\textit{et al.}}: Bare Demo of IEEEtran.cls for Journals}

\maketitle

\begin{abstract}
Models of the cortico-basal ganglia network and volume conductor models of the brain can provide insight into the mechanisms of action of deep brain stimulation (DBS). In this study, the coupling of a network model, under parkinsonian conditions, to the extracellular field distribution obtained from a three dimensional finite element model of a rodent's brain during DBS is presented. This coupled model is used to investigate the influence of uncertainty in the electrical properties of brain tissue and encapsulation tissue, formed around the electrode after implantation, on the suppression of oscillatory neural activity during DBS. The resulting uncertainty in this effect of DBS on the network activity is quantified using a computationally efficient and non-intrusive stochastic approach based on the generalized Polynomial Chaos. The results suggest that variations in the electrical properties of brain tissue may have a substantial influence on the level of suppression of oscillatory activity during DBS. Applying a global sensitivity analysis on the suppression of the simulated oscillatory activity showed that the influence of uncertainty in the electrical properties of the encapsulation tissue had only a minor influence, in agreement with previous experimental and computational studies investigating the mechanisms of current-controlled DBS in the literature.
\end{abstract}

\begin{IEEEkeywords}
Deep brain stimulation, Uncertainty quantification, Finite element method, Neural network models.
\end{IEEEkeywords}

\IEEEpeerreviewmaketitle

\section{Introduction}
\IEEEPARstart{D}{eep brain stimulation} (DBS) is a neurosurgical therapy used to treat symptoms of neurodegenerative disorders including Parkinson's disease (PD). Despite its approval for the treatment of PD and its widely acknowledged success, the mechanisms of action of DBS remain uncertain. Electrophysiological studies in patients and in animal models have shown that PD is associated with excessively synchronized oscillatory neural activity in the cortico-basal ganglia network accompanied by an increased power in the beta frequency range (\SI{15}{Hz}-\SI{35}{Hz}) \cite{litvak2011, eusebio2008, kuhn2008, bronte2009}.\newline \indent
For almost two decades, computational models of DBS have been developed to investigate its mechanisms of action. Among these, biological models of the cortico-basal ganglia network, based on the excitatory and inhibitory connections between the different deep brain nuclei, have been used to investigate the effect of square-wave DBS signals on the network dynamics \cite{rubin2004, hahn2010, kang2013}. In these network models, the DBS signal was applied as an intracellular current to the neurons of the target nuclei, typically the subthalamic nucleus (STN) or globus pallidus interna (GPi) for PD DBS. \rev{This modeling approach does not capture the reduction in amplitude of the extracellular potential with increasing distance from the electrode or the effect of the extracellular field distribution along the neuron axons. The distribution of the electric field is determined by the electrode geometry, the electrode-tissue-interface, and the heterogeneity of the brain tissue as well as its electrical properties.}\newline \indent
Computational studies based on volume conductor models of the brain have shown that neural activation in the proximity of the stimulation electrode depends strongly on the field distribution caused by the DBS signal and, therefore, on the aforementioned factors \cite{butson2005, grant2010, schmidt2013biostec}. The electrical properties of biological tissue such as brain tissue and encapsulation tissue are subject to uncertainty due to wide variations in the values reported in the literature, most of which is  based upon measurement \textit{ex vivo} \cite{gabriel2009}. \rev{There is considerable variation among reported conductivity and relative permittivity values at lower frequencies in particular. This is due largely to the limitations of available techniques for measuring dielectric properties of biological tissues within the low frequency range, including electrode polarization \cite{gabriel2009}.}\newline \indent
Simulation results suggest that the uncertainty in the electrical properties of brain tissue \rev{can have} a substantial influence on the field distribution and stimulation amplitude required to activate the target nuclei \cite{schmidt2013}. The impact that the volume conductor effects and material property uncertainty \rev{consequently have} on the network dynamics and the level of suppression of oscillatory neural activity for a given stimulus, however, is not clear.\newline \indent
To investigate the influence of parametric uncertainties in the electrical properties of brain tissue on the suppression of beta activity in the cortico-basal ganglia network, it is necessary to quantify the uncertainty in the time-dependent extracellular potential distribution in the brain as well as in the firing patterns of the neuronal network. These kind of stochastic problems can be solved with standard brute-force approaches such as Monte-Carlo simulation \rev{This approach, however, requires} a large number of subsequent evaluations of the computational model. Alternatively, computationally efficient approaches such as that based on the generalized Polynomial Chaos (gPC) technique may be used. This approach substantially reduces the number of model evaluations required by generating a surrogate model of the quantities of interest that may be more easily evaluated. \rev{The reduction of the number of required model evaluations for generating this surrogate model is carried out by a multi-variate interpolation, which requires the model to be evaluated only for specific samples of the brain tissue electrical properties to compute the multi-variate interpolation polynomials.}\newline \indent
\rev{The aim of this study was to explore how uncertainty in the electrical properties of brain tissue can effect predicted levels of suppression of pathological neural oscillations in the cortico-basal ganglia network during DBS. To address this, a coupled model of volume conduction within a rodent's brain and neural activity within the cortico-basal ganglia network during DBS was developed.} The gPC method was then applied to quantify the influence of uncertainty in the electrical properties of brain tissue and encapsulation tissue. The network model is based on a model of the cortico-basal ganglia,  which was previously used to investigate the relative effects of antidromic and orthodromic activation of cortico-STN afferents \cite{kang2014}. To enable coupling between the field distribution in the rodent's brain and the network model, the nodes of the cortico-STN afferent axons were first identified to allow application of the corresponding extracellular potential. The coupled model was then expanded to a stochastic model with the electrical properties of the brain and encapsulation tissue modeled as random variables. This expanded model was used to investigate the influence of varying tissue electrical properties on the suppression of beta-band oscillatory activity in the network. A preliminary version of this study has been reported in \cite{schmidt2015ner}.

\section{Methods}

\subsection{Cortico-Basal Ganglia Network Model}
\begin{figure}[t]
    \centerline{\psfig{figure=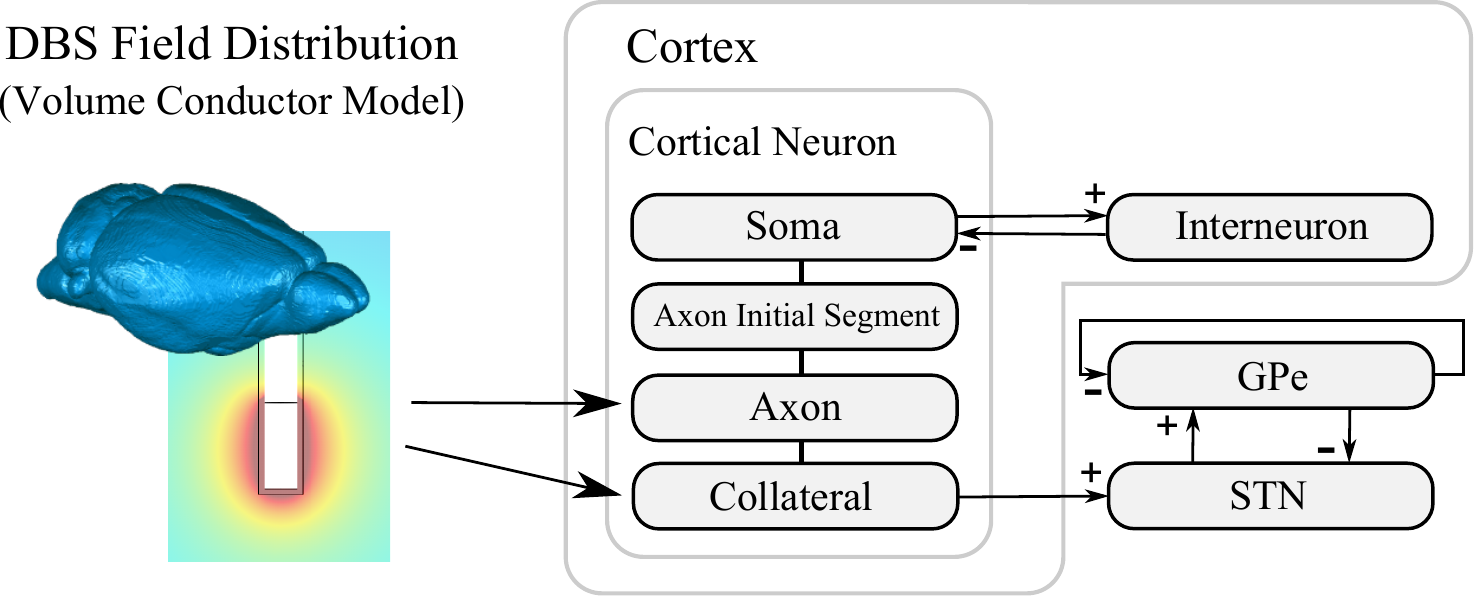,width=1.0\columnwidth} }
    \caption{Schematic illustration of the cortico-basal ganglia network with the DBS applied to the cortical axons and branching collaterals to the STN. Excitatory pathways and inhibitory pathways are indicated by "+" and "-" signs.}
    \label{fig:bgmodel}
\end{figure}

The network model is based on a previous cortico-basal ganglia model in which the stimulus was applied to cortico-STN afferent axons using an analytical, purely resistive point source to model the field distribution during DBS \cite{kang2014}. In this network model, the neurons in the STN, and globus pallidus external (GPe) are modeled as single compartment, conductance based Hodgkin-Huxley type models, with threshold crossing models representing the interneurons within the cortex. \rev{The STN was based on a parkinsonian rat model which captures the generation of plateau potentials \cite{otsuka2004}. The cortex was composed of a threshold crossing model representing cortical interneurons \cite{izhikevich2003} and a multi-compartment model of cortical neurons consisting of a soma, an axon initial segment, a myelinated main axon, comprising five nodes of Ranvier and four internodes, and an axon collateral which connects to the STN. The soma was based on the regular spiking neuron model developed by Pospischill et al \cite{pospischil2008}, while the axon initial segment, axon, and collateral were based on a mouse model developed by Foust et al \cite{foust2011}.} Each nucleus and neuron type consisted of a population of $30$ neurons. The nuclei were connected by excitatory (glutamergic) and inhibitory (GABA) synapses (Fig. \ref{fig:bgmodel}). \rev{The type and direction of connections between nuclei of the cortico-basal ganglia were based on data available for the network in the diseased state \cite{shepherd2013}, however, it is more difficult to ascertain the exact number of connections between individual neurons in different nuclei. The STN neurons received inhibitory input from the GPe and excitatory input from three randomly chosen cortical neurons. The PD state of the network was simulated by applying a DC current to the cortical neurons, as the STN receives substantial direct cortical input \cite{albin1989}. The model connections and parameters were further designed to match antidromic firing rate results from a parkinsonian rat model \cite{li2012}, which showed an increasing frequency of antidromic spikes as the frequency of stimulation was increased, until reaching a peak at \SI{125}{Hz} stimulation frequency.} and increasing the synaptic gains within the network as described in \cite{dunn2013}, leading to oscillatory \rev{activity in the} STN neurons within the beta frequency range. The network model was implemented in NEURON~7.1 \cite{hines2003}. A time step of \SI{1}{\micro s} was used and the random seed was reused to ensure same connections between the nuclei and synaptic noises for each computation. \rev{The model was initialized for 5 seconds to reach the PD steady-state} before the application of DBS. DBS was then modeled by applying the computed extracellular potential for 2 seconds to both the axon collateral and the main axon of the cortical neurons. Pulse trains representing the firing times of all STN neurons were summed and the power spectrum of the resulting composite pulse train was computed. \rev{The level of beta activity within the STN was then estimated as the integral of the power spectrum of the composite spike train in the frequency range between \SI{15}{Hz} and \SI{30}{Hz} \cite{brown2000}. The power spectra and level of beta activity in the GPe and cortical neurons were similarly estimated.}

\subsection{Extracellular Field Distribution in the Rodent's Brain}
\begin{figure}[t]
    \centerline{\psfig{figure=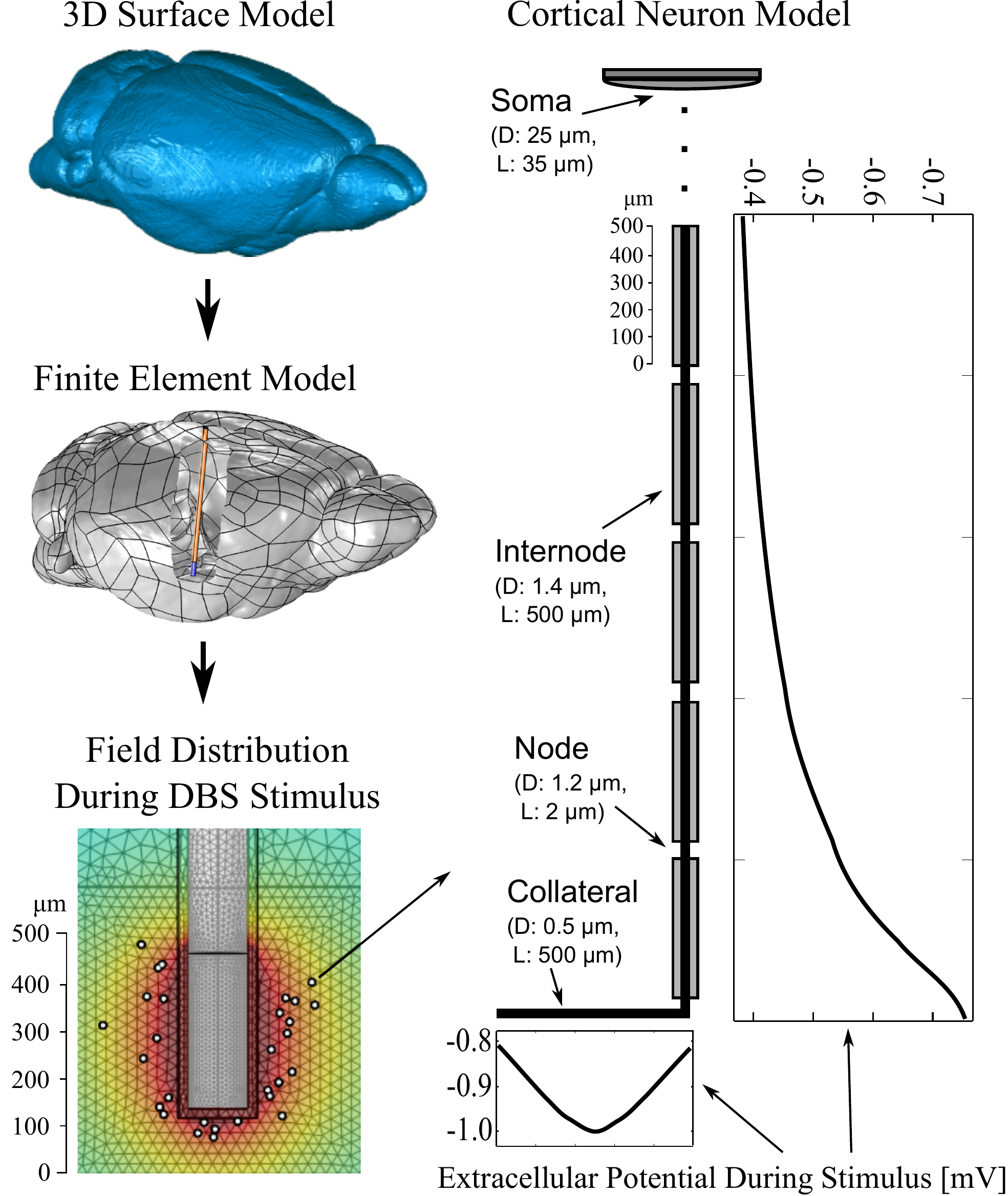,width=1.0\columnwidth} }
    \caption{\rev{Schematic illustration of the coupling between the volume conductor model and the cortical neurons. The collaterals are positioned randomly in the proximity of the electrode contact. The time-dependent extracellular potential, shown here for a stimulation amplitude of \SI{150}{\micro A} is applied along the collateral and the nodes and internodes along the axon.}}
    \label{fig:model}
\end{figure}
The volume conductor model of the rodent's brain is based on a three-dimensional digital atlas database of the adult C57BL/6J mouse brain. The segmented brain atlas data comprises an average from magnetic resonance microimages of 10 adult males and has an isotropic resolution of \SI{47}{\micro m} \cite{ma2005}. The brain surface was parametrized using non-uniform rational B-splines with the software Geomagic Studio\textregistered \, (http://www.geomagic.com). The finite element model of the rodent's brain was generated using Comsol Multiphysics\texttrademark \, (http://www.comsol.com) by importing the parametrized brain surface model and a model of a DBS microelectrode (diameter:  \SI{125}{\micro m}, electrode contact height: \SI{320}{\micro m} \cite{dehaas2012}), as illustrated in Fig. \ref{fig:model}. The microelectrode was positioned in the STN area ventral and lateral to the thalamus by comparing the image data with the Allen Brain Atlas of the mouse brain (http://atlas.brain-map.org). The electrical ground of the model was set to the area of the cerebellum to model the placement of the ground electrode in the neck of the mouse. A square-wave current-controlled DBS signal with a frequency of \SI{130}{Hz} and a pulse duration of \SI{60}{\micro s} was applied to the electrode contact via a Dirichlet boundary condition \cite{gimsa2005}. The time-dependent field distribution was computed using the Fourier Finite Element Method \cite{butson2005} by solving the Laplace equation
	\begin{align}
		\nabla\kappa(\boldsymbol r)\nabla\varphi(\boldsymbol r) = 0
		\label{eq:qs}
	\end{align}
where  $\kappa(\boldsymbol r)$ is the electric conductivity and $\varphi(\boldsymbol r)$ the electric potential at location $\boldsymbol r$ in the computational domain $\Omega$. Due to reactions of the body to the implant, scar tissue forms an encapsulation layer around the electrode. The electrical properties of this highly resistive layer as well as brain tissue vary considerably within the literature \cite{grant2010, yousif2009}. To capture these variations, their electrical properties were modeled as uniformly distributed random variables with a standard deviation of \SI{10}{\percent}, which corresponds to a relative deviation with respect of the 97.5 percentile and 2.5 percentile of \SI{16.5}{\percent} (Table \ref{tab:tissues}). The mean value of brain tissue conductivity was chosen according to the value of white matter at a frequency of \SI{2}{kHz} \cite{gabriel1996}, which constitutes a good approximation of the dispersive nature of the electrical properties in volume conductor models of the brain \cite{schmidt2013biostec}. The mean value of the encapsulation tissue was set to half the mean value of the brain tissue \cite{schmidt2013}.\newline \indent
\rev{DBS was applied to the cortical collaterals and axons, following the results of an optogenetic study in 6-OHDA rats that found that theraputic effects of STN DBS could be accounted for by direct selective stimlation of afferent fibers projecting from the cortex to the STN \cite{gradinaru2009}.} The collaterals of the cortical neurons branching to the STN were randomly distributed within a distance of \SI{250}{\micro m} from the center of the electrode contact using uniformly distributed random variables for their \rev{Cartesian} coordinates (Fig. \ref{fig:model}). The collaterals were orientated perpendicular to the main axis of the electrode, with the longitudinal axes of each collateral centered at the electrode. The axons of the cortical neurons, from which the collaterals branched, were orientated perpendicular to the collaterals and parallel to the electrode. To ensure a sufficient resolution of the finite element mesh with respect to the size of the collaterals and the axons, manual mesh refinement was applied by setting a maximum mesh size of \SI{25}{\micro m} in the area of the collaterals, \SI{50}{\micro m} in the area of the axons, and \SI{10}{\micro m} at the electrode contact surface. The final mesh consisted of approximately $1.2$ million elements resulting in approximately $1.7$ million degrees of freedom using quadratic basis functions. The linear system was solved using generalized minimal residual method with a relative tolerance of $1\cdot10^{-6}$.
\begin{table}[!t]
	\centering
  \renewcommand{\arraystretch}{1.3}
	\caption{\label{tab:tissues}Lower and upper boundary, mean value, and relative deviation of the uniformly distributed conductivities of brain tissue and encapsulation tissue.}
	\begin{tabular}{lccccc}
			\hline
			\bfseries Tissue type & \multicolumn{3}{c}{\bfseries Conductivity $\sigma$ $\left[\mathrm{mS/m}\right]$} & \\
			& \bfseries Min & \bfseries Mean & \bfseries Max & \bfseries Rel. Deviation \\
			\hline \hline
			Brain tissue & $64.0$ & $77.4$ & $90.8$ & \SI{16.5}{\percent} \\
			Encapsulation tissue & $32.0$ & $38.7$ & $45.4$ & \SI{16.5}{\percent}\\
			\hline
	\end{tabular}	
\end{table}

\subsection{Coupling of the Extracellular Field Distribution to the Network Model}
\label{subsec:coupling}
\rev{The spatial distribution of the cortical neurons resulted in differences in the extracellular potential along the collaterals and main axons of each neuron. To incorporate the effect of variations in extracellular field distribution on the dynamics of the neural network, the extracellular field was coupled to the nodes of the target axons in the discretized network model. Following the approach of previous computational studies to couple the field distribution from a finite element model to neuron models \cite{butson2005, astrom2015}, the time-dependent potential of the finite element model was computed at the location of each segment of the cortical axons and collaterals in the neural network model and was applied as a discrete time-dependent extracellular potential.\newline\indent} In a previous study, the extracellular potential during DBS was assumed to preferentially activate the branching collaterals and the extracellular potential was, therefore, applied solely at the collateral \cite{kang2014}. \rev{To capture potential effects of the extracellular potential on both the cortical axons and collaterals, in the present study the time-dependent extracellular potential computed using the volume conductor model was applied to each segment along both collaterals and axons at a total of $10$ equidistant nodes along each collateral and $55$ nodes along each axon.}

\subsection{Uncertainty Quantification Using Polynomial Chaos}
During the application of DBS, the firing times of neurons within the cortex and STN will vary with the time-dependent extracellular potential. The beta-band activity of the neural network model will thus depend on the time-dependent extracellular potential distribution computed using the volume conductor model. The field distribution, in turn, depends on the conductivity of the encapsulation tissue and brain tissue. In volume conductor models, the isotropic conductivity, and permittivity if considered, of the biological \rev{tissues} are typically modeled using a single value. There is, however, considerable uncertainty in the true values \rev{of the dielectric properties}, due to challenges in the measurement of the electrical properties of biological tissue, especially in the low \SI{}{Hz} and \SI{}{kHz} frequency regime \cite{gabriel2009}. To capture the uncertainty arising from this variation, in this study, the conductivity of each tissue type was described using a random variable to model its uncertainty. To compute the statistics of the quantities of interest, standard stochastic methods such as Monte-Carlo simulation could be used. To provide a sufficient accuracy in the stochastic measures, Monte-Carlo simulation requires a large number of random samples of the model parameters \cite{xiu2010}. For each sample, the whole computational model \rev{must} be solved from which a corresponding sample of each quantity of interest can be computed. If the deterministic model is computationally expensive, as it is the case in the present study, this approach becomes computationally inefficient.\newline \indent
To reduce the computational effort for determining the stochastic measures of the quantities of interest, a non-intrusive variant of the gPC technqiue is applied. This technique determines a surrogate model of each quantity of interest by expanding them in a series of multi-variate orthogonal polynomials and computing the model solution at predefined nodes in the parameter space\rev{, which is spanned in this study by the uncertain electrical properties of brain tissue} \cite{constantine2012}. These nodes are determined by combining the one-dimensional Clenshaw-Curtis quadrature rule with a combinatorical algorithm to generate a multi-dimensional sparse grid representation of the parameter space (sparse-pseudospectral approach). These model solutions at the grid nodes are used to compute the coefficients of the multi-variate polynomial series expansion with multi-dimensional numerical integration. The number of grid nodes depend on the polynomial degree $p$ of the gPC expansion as well as on the number of model parameters $M$ \cite{nobile2008}. In general, gPC is computationally advantageous over Monte-Carlo simulation for a small number of model parameters (e.g. $M=6$ and $p=3$ requires $389$ evaluations of the model). In addition, the series expansions of the quantities of interest provided by the gPC approach allow for a simple global sensitivity analysis by Sobol' indices. These indices determine the sensitivity of these quantities with respect to each uncertain parameter as well as their stochastic interactions by evaluating their conditional variances \cite{sobol2001}. In this relation, the term conditional variance describes the variance of the quantity of interest with respect to each uncertain parameter, \rev{the uncertain electrical properties of brain tissue, or any combination of them.}\newline \indent
The gPC and sparse-pseudo-spectral approach has previously been successfully applied to other bio-electrical applications including optimizing the stimulation protocol in a multi-electrode transcranial direct current stimulation setup as well as for an electro-stimulative hip-revision system \cite{schmidt2015jne,schmidt2015jbhi}.\newline \indent
\begin{figure*}[t]
    \centerline{\psfig{figure=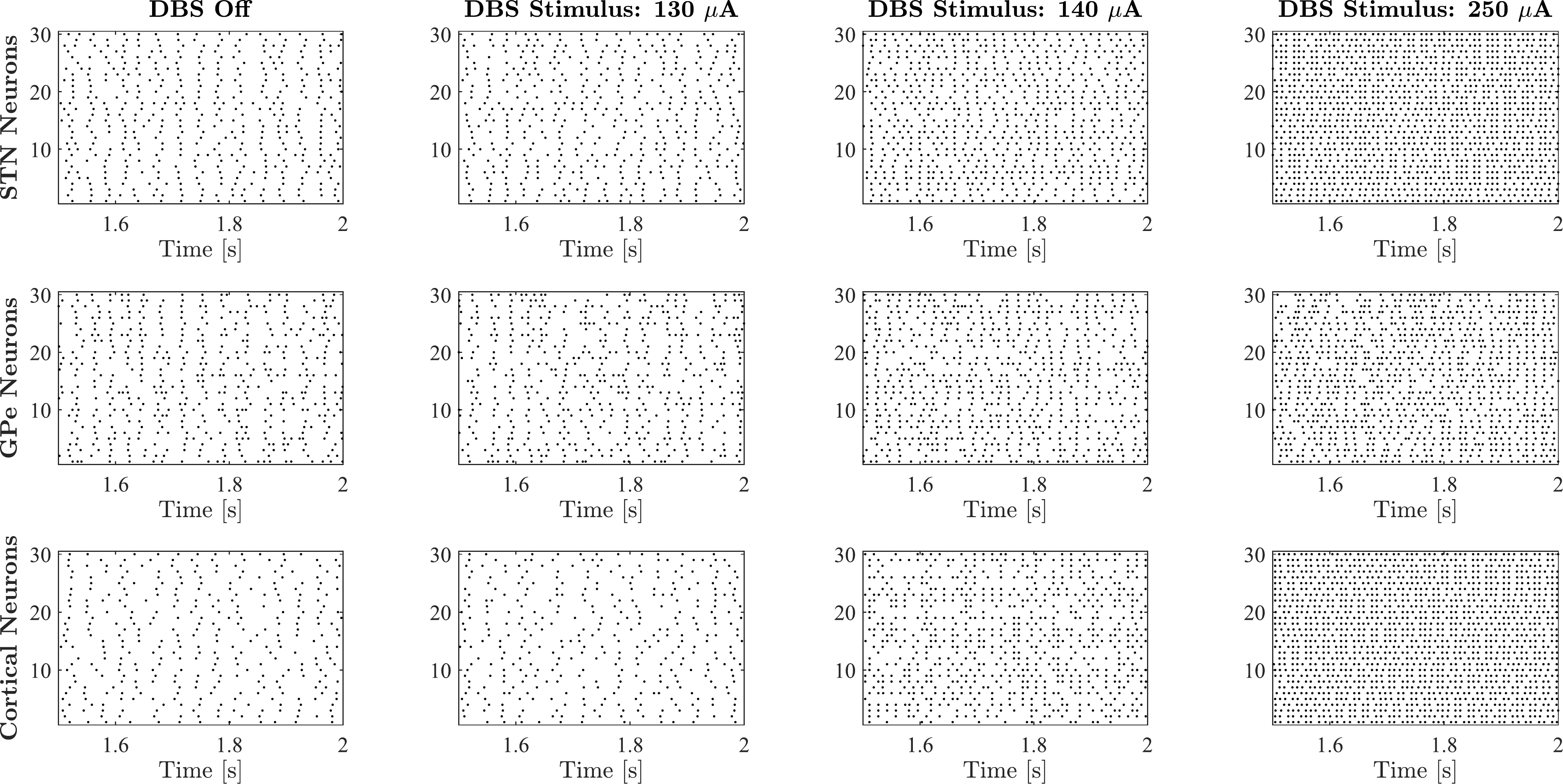,width=1.0\textwidth} }
    \caption{\rev{Firing patterns of the STN, GPe, and Cortical Neurons in the PD state with no DBS applied, and for DBS stimulation amplitudes of \SI{130}{\micro A}, \SI{140}{\micro A}, and \SI{250}{\micro A}. The patterns were recorded for the cell bodies of each neuron.}}
    \label{fig:firing_patterns}
\end{figure*}
\begin{figure*}[t]
    \centerline{\psfig{figure=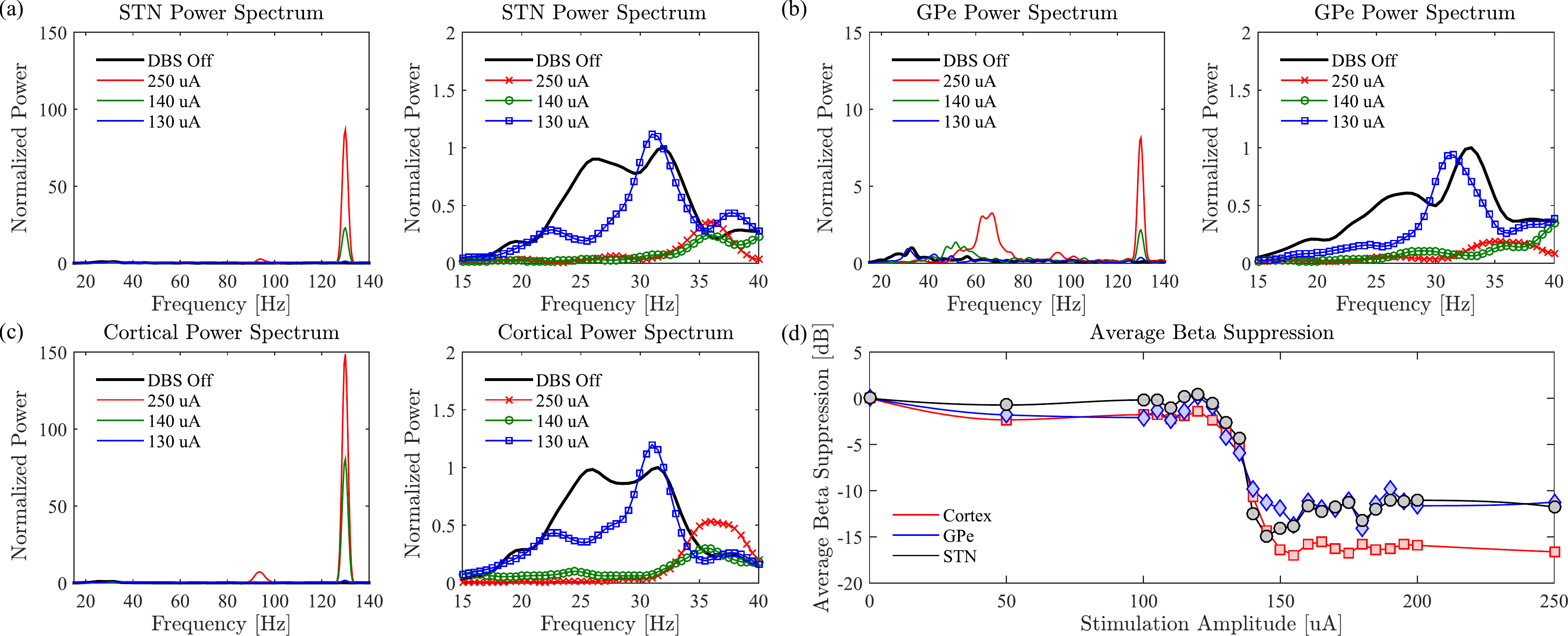,width=1.0\textwidth} }
    \caption{\rev{Power spectral densities, normalized with respect to the DBS Off condition, of the (a) STN, (b) GPe, and (c) cortical neurons for different stimulation amplitudes. The average beta suppression is shown for all three nuclei in (d).}}
    \label{fig:power_spectra_mean}
\end{figure*}
The method was implemented in Matlab as \rev{a} function package, which determines the deterministic parameter samples on the chosen sparse grid based on the predefined properties of the random model parameters and the polynomial degree $p$, as described in \cite{schmidt2015jne}. The list of parameter samples was used to subsequently initialize the solution process of the volume conductor model with COMSOL and of the network model with NEURON. From the resulting list of solutions, the corresponding sample list of the desired quantities of interest \rev{including} the extracellular potential and the level of beta suppression was derived, from which the respective series expansion, as mentioned above, were computed. The evaluation of these series expansions of the quantities of interest for any samples in the parameter space require only the subsequent evaluation of the product of basis polynomials, which provides a suitable model for Monte Carlo simulation \cite{nobile2008}. Therefore, the determined surrogate models of the quantities of interest allow for the application of Monte Carlo simulation with a large number of random samples to provide sufficiently accurate stochastic measures. 

\section{Results}

\begin{figure*}[t]
    \centerline{\psfig{figure=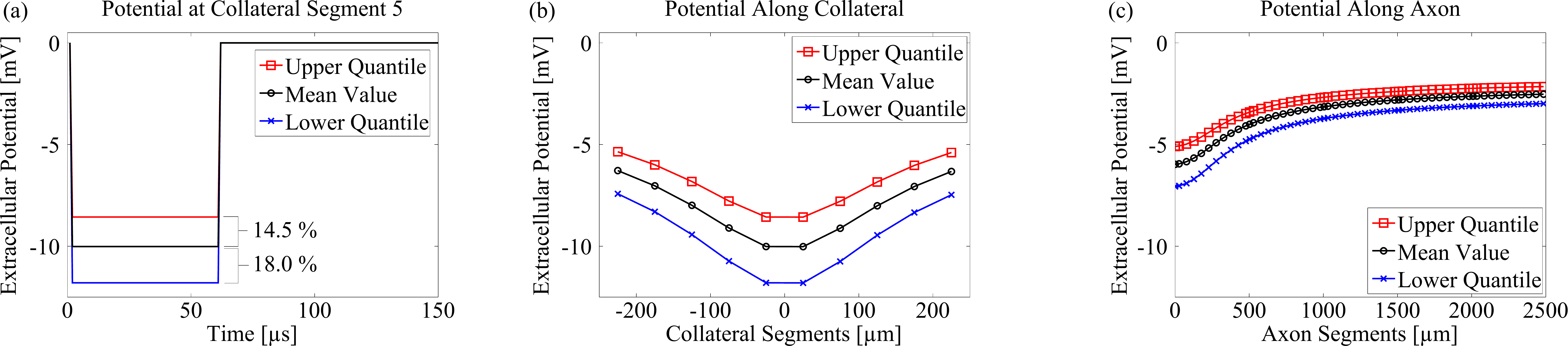,width=1.0\textwidth} }
		\caption{Mean value, lower (2.5 percentile), and upper quantile (97.5 percentile) of the uncertain extracellular potential (a) at the 5th segment of a reference cortical collateral positioned perpendicular to the electrode contact at a radial distance of \SI{100}{\micro m}, (b) along the cortical collateral at at time \SI{50}{\micro s}, (c) and along the corresponding cortical axon.}
    \label{fig:uq_epot}
\end{figure*}

\begin{figure}[t]
    \centerline{\psfig{figure=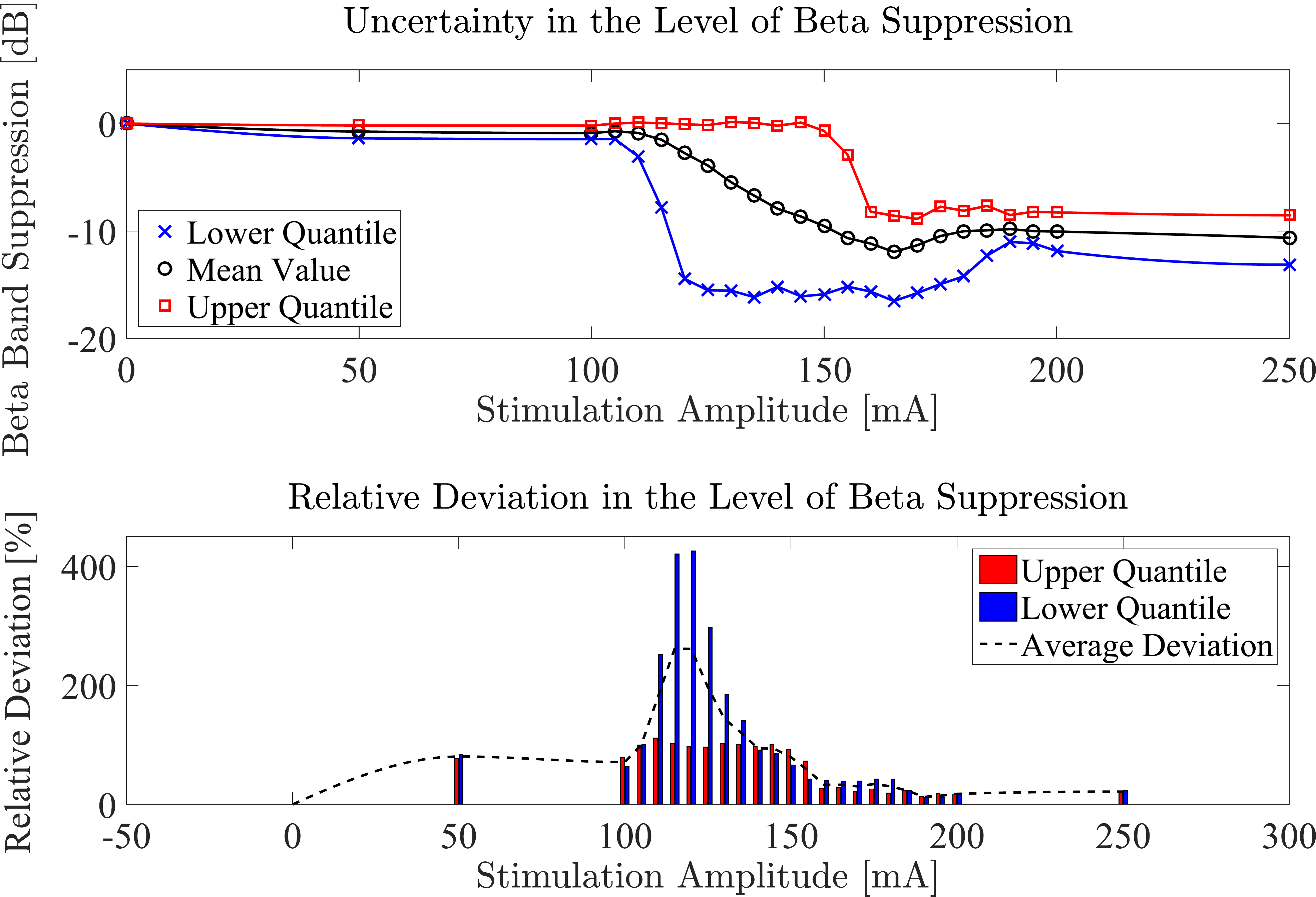,width=1.0\columnwidth} }
	\caption{\rev{(top) Mean value, lower quantile (2.5 percentile), and upper quantile (97.5 percentile) of the beta-band suppression across all STN neurons with increasing stimulation amplitude. (bottom) Relative deviation between the mean value of the beta-band suppression and the corresponding lower quantile (blue) and upper quantile (red).}}
	\label{fig:stim_uq_bars}
\end{figure}

Under simulated parkinsonian conditions, STN neurons in the model exhibited synchronized activity within the beta frequency range (Fig. \ref{fig:firing_patterns}). To investigate the response of the network model to DBS of increasing stimulation intensity, the level of beta suppression in the STN neurons within the network model was investigated for stimulation amplitudes between \SI{50}{\micro A} and \SI{250}{\micro A}. \rev{At} each stimulation amplitude, the corresponding extracellular field distribution was computed using the mean conductivity values of encapsulation and brain tissue, as \rev{given} in Table \ref{tab:tissues}. The beta-band power of \rev{the point process} representing the firing times of the STN neurons progressively decreased as the amplitude of the DBS signal was increased (Fig. \ref{fig:power_spectra_mean}(d)).
As the DBS amplitude was increased, an \rev{increase in the} number of activated collaterals and axons was observed\rev{, leading to an increase in the region of directly stimulated tissue.} (Fig. \ref{fig:firing_patterns}). The collateral and axonal activation orthodromically activated the STN \rev{leading} to a gradual entrainment of the STN neurons to the stimulus pulse and subsequent reduction of the level of beta-band oscillations \rev{(Fig. \ref{fig:power_spectra_mean}(a)), which is also noticeable for the GPe neurons (Fig. \ref{fig:power_spectra_mean}(b)). The cortical neurons and inhibitory interneurons were simultaneously antidromically activated, reducing beta-band activity from the cortex to the STN (Fig. \ref{fig:power_spectra_mean}(c)).} For stimulation amplitudes of \SI{150}{\micro A} and higher, a slight increase in the amplitude of the beta-band high-frequencies between \SI{30}{Hz} and \SI{40}{Hz} can be seen. The increased activity in this frequency range is accompanied by a more dense spiking pattern at \rev{\SI{250}{\micro A} compared to that at \SI{150}{\micro A} (Fig. \ref{fig:firing_patterns}). \rev{Nevertheless, a clear suppression of most of the power spectrum due to an almost complete entrainment of all STN, GPe, and cortical }neurons, is observed for all stimulation amplitudes between \SI{140}{\micro A} and \SI{250}{\micro A} \rev{(Fig. \ref{fig:power_spectra_mean}(d))}.}\newline \indent
\rev{Following evaluation of the network response with the tissue conductivities in the volume conductor model set to their respective mean values, the gPC method with a polynomial order of $p=4$ was used to investigate the influence of uncertainty in the tissue conductivity on the quantities of interest.} Using the sparse-pseudo spectral approach and the multi-dimensional implementation of the numerical integration method based on the Clenshaw-Curtis rule, the computation of the corresponding gPC series expansions required the evaluation of the computational model for $29$ deterministic samples of the tissue conductivity \rev{for the computation of the uncertain extracellular potential}. The uncertainty of the extracellular potential within the proximity of the stimulation electrode was approximately \SI{15}{\percent} and \SI{18}{\percent} with respect to the upper quantile (97.5 percentile) and the lower quantile (2.5 percentile), respectively. Exemplarily results are presented for a reference cortical collateral positioned perpendicular to the electrode at a radial distance of \SI{100}{\micro m} and for a stimulation amplitude of \SI{250}{\micro A} in Figure \ref{fig:uq_epot}(a). The slightly larger deviation with respect to the lower quantile compared to the upper quantile arises from the inversely proportional relationship between conductivity and potential distribution for current-controlled stimulation, which results in an asymmetric probability density function and, therefore, unequal differences \rev{in} the mean value \cite{schmidt2013}. The level of uncertainty in the time-dependent extracellular potential at a specific location was similar at the other locations of the collateral and axon segments (Fig. \ref{fig:uq_epot}(b,c)). Comparing the average relative deviation of approximately \SI{16.3}{\percent} in the extracellular potential to the uncertainty in the tissue conductivities of approximately \SI{16.5}{\percent} (Table \ref{tab:tissues}), an almost 1:1 relation between the level of their uncertainty can be noted. \rev{A similar relationship was observed} for the uncertainty in the bulk tissue impedance of the volume conductor model (Table \ref{tab:uq_Z}). An increase of the conductivity of brain tissue as well as encapsulation tissue resulted in a decrease in the bulk tissue impedance as well as the extracellular potential distribution, and vice versa. Regarding the accuracy of the stochastic measures in these quantities of the volume conductor model, a doubling in the polynomial order of $p=8$, which required the computation of additional $36$ model evaluations, resulted in a difference of below $9\cdot10^{-5}\,\SI{}{\percent}$ and \SI{0.03}{\percent} in their mean values and relative deviations, respectively. Since the extracellular potential distribution within the volume conductor model scales \rev{linearly} with respect to the applied stimulation amplitude due to the linear material properties and the Laplace equation (\ref{eq:qs}), the solution for different stimulation amplitudes was obtained by simply multiplying the reference solution by the corresponding stimulation amplitude.\newline \indent
\begin{table}[!t]
	\centering
  \renewcommand{\arraystretch}{1.3}
	\caption{\label{tab:uq_Z}Mean value, lower quantile (2.5 percentile), and upper quantile (97.5 percentile) of the uncertain bulk tissue impedance determined from the volume conductor model with uncertain tissue properties.} 
	\begin{tabular}{lcc}
			\hline
			\bfseries Stochastic Measure & \bfseries Impedance [$\mathsf{k\Omega}$] & \bfseries Rel. Deviation \\
			\hline \hline
			Lower quantile & $15.9$ & $\SI{15.4}{\percent}$\\
			Mean value & $13.8$ & $0.0$\\
			Upper quantile & $12.0$ & $\SI{13.1}{\percent}$\\
			\hline
	\end{tabular}	
\end{table}
While this linear property of the volume conductor model allowed for a reuse of the reference solution, which only had to be computed once, the deterministic solution of the network model had to be reevaluated for each stimulation amplitude due to the nonlinearities in the dynamics of the network model and individual neurons \cite{kang2014}. This nonlinearity in the response of the network model with respect to the stimulation amplitude is apparent in the suppression-amplitude relation shown in Figure \rev{\ref{fig:power_spectra_mean}. Recent studies have demonstrated suppression of beta band activity witihn the basal ganglia local field potential (LFP) during DBS, accompanied by a reduction in the parkinsonian motor symptoms of rigidity and bradykensia \cite{kuhn2009}. The presence of beta band LFP oscillations are interpreted as being indicative of pathological synchronization between neurons and have attracted substantial interest as a potential biomarker of the parkinsonian state \cite{hammond2007}. The influence of uncertainty in the tissue electrical properties on the level of suppression of beta frequency neural oscillations in the STN was thus chosen for investigation. The effect on uncertainty in the material properties of the brain and encapsulation tissue on uncertainty in the level of suppression of synchronous beta band activity among the STN neurons is shown in Figure \ref{fig:stim_uq_bars}.} For lower conductivity values, as in the case of the lower quantile values, a decrease in the STN beta-band activity occurred at lower stimulation amplitudes as the potential within the bulk tissue for a given stimulus amplitude was higher. Higher stimulation amplitudes were \rev{correspondingly} required to elicit comparable levels of beta suppression when higher conductivity values, as in the case of the upper quantile values, were assumed within the model. Therefore, the uncertainty in the tissue conductivities introduced a large uncertain area in the window of effective stimulation spanned by the lower and upper quantiles (Fig. \ref{fig:stim_uq_bars}(a)).
 Stimulation amplitudes below this window were too low to sufficiently modify the membrane potential along the cortical axons and collaterals to have an effect on the spiking pattern in the STN neurons. Conversely, stimulation amplitudes above this window seem to have no additional effect on the spiking pattern. \rev{This effect} can be explained by investigating the membrane potential along the cortical neurons. These neurons were already entrained by the DBS stimulus and, therefore, a further increase in the stimulation amplitude did not influence the dynamics of the already stimulated cortical \rev{axons (Fig. \ref{fig:firing_patterns}}). The resulting level of uncertainty was greatest for a stimulation amplitude of \SI{120}{\micro A} with an average value of approximately \SI{260}{\percent} (Fig. \ref{fig:stim_uq_bars}). The asymmetry in the relative deviation with respect to the upper and lower quantile of the extracellular potential and bulk tissue impedance (Fig. \ref{fig:uq_epot} and Table \ref{tab:uq_Z}) is also noticeable in the uncertainty of the level of beta suppression. \rev{In general,} the uncertainty in the level of beta suppression in the network model was substantially larger than the predefined uncertainty in the tissue conductivities, which can be ascribed to the nonlinearity in the mathematical formulation of the network model. This nonlinear behavior affects also the accuracy of the stochastic measures with respect to the polynomial order of the gPC. \rev{The uncertain beta supression of beta activity in the network model was computed using gPC with a polynomial order of $p=8$, which required the evaluation of $65$ network models for each stimulation amplitude.} \rev{A doubling of the polynomial order of $p=16$ resulted in a difference of less than \SI{0.1}{\percent} and \SI{9.1}{\percent} in the mean value and relative deviations of the level of beta suppression, respectively, evaluated at a stimulation amplitude of \SI{120}{\micro A}, which is substantially larger than for the stochastic measures of the volume conductor model. Nevertheless, the accuracy in the stochastic measures of the network model is sufficiently large for an investigation of the overall trend of its uncertainty.} \rev{Regarding the required number of random samples to determine the stochastic measures of the surrogate models obtained by the gPC method, a number of \SI{100.000} random samples was sufficient to provide an accuracy of approximately \SI{0.1}{\percent}}.\newline \indent
\begin{figure}[t]
    \centerline{\psfig{figure=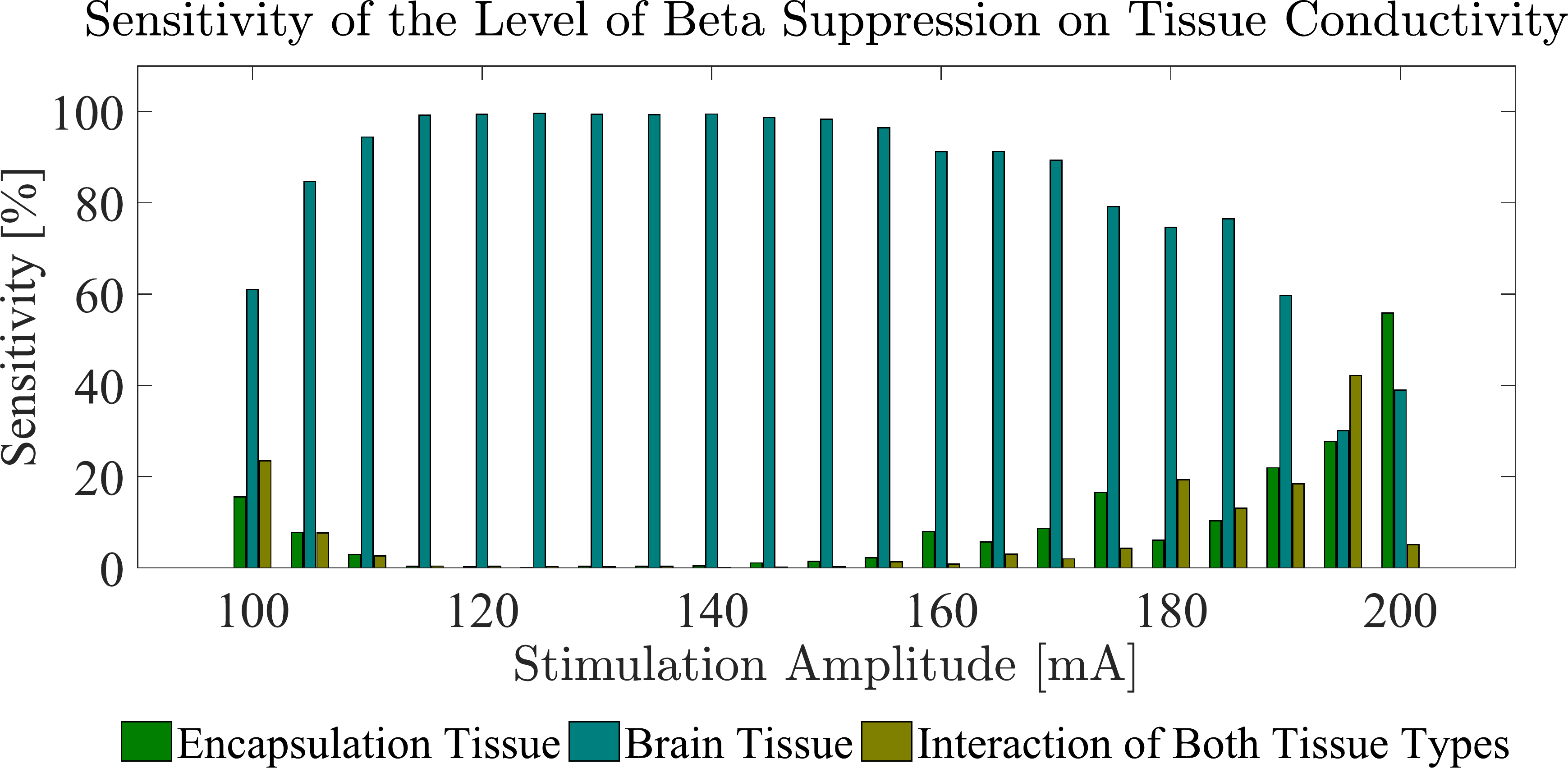,width=1.0\columnwidth} }
	\caption{\rev{Sensitivity of the beta-band suppression to the uncertainty in brain tissue, encapsulation tissue, and the interaction of the uncertainty in both tissue types.}}
	\label{fig:sobol}
\end{figure}
\begin{table}[!t]
	\centering
  \renewcommand{\arraystretch}{1.3}
	\caption{\label{tab:sobol}Sensitivity of the bulk tissue impedance and extracellular potential with respect to the uncertainty in brain $\sigma_\mathsf{brain}$ and encapsulation tissue $\sigma_\mathsf{encap}$ as well as their interaction.} 
	\begin{tabular}{lccc}
			\hline
			\bfseries Quantities of Interest & \multicolumn{3}{c}{\bfseries Sensitivity [\SI{}{\percent}]} \\
			& \bfseries $\sigma_\mathsf{brain}$ & \bfseries $\sigma_\mathsf{encap}$ & \bfseries $(\sigma_\mathsf{brain},\sigma_\mathsf{encap})$ \\
			\hline \hline
			Bulk Tissue Impedance & $\mathbf{91.1}$ & $8.9$ & $0.0$\\
			Extracellular Potential & $\mathbf{99.9}$ & $0.1$ & $0.0$\\
			\hline
	\end{tabular}	
\end{table}
The investigation of the sensitivity of the quantities of interest by evaluating the Sobol' indices revealed that the major influence on the uncertainty in the extracellular potential and the bulk tissue impedance is due to the uncertainty in brain tissue conductivity (Table \ref{tab:sobol}). As expected, due to the linearity of the volume conductor model, no interaction effects of the uncertainty in both tissue conductivities is noticeable. Since the dynamics of the network model depend on the extracellular potential along the axons and collaterals of the cortical neurons distributed in the volume conductor model, a similar behavior would be assumed for the sensitivity in the level of beta suppression. Investigating this sensitivity for stimulation amplitudes between \SI{120}{\micro A} and \SI{170}{\micro A}, this \rev{is confirmed} (Fig. \ref{fig:sobol}). Nevertheless, the influence of uncertainty in encapsulation tissue conductivity as well as its interaction with the uncertainty in brain tissue conductivity seem to increase for stimulation amplitudes below and above that range, which is assumed to be caused by dynamic effects related to the random noise in the network model and high sensitivity of the neurons to the extracellular potential when close to the threshold for excitation.

\section{Discussion}
The coupling of a cortico-basal ganglia network model with the extracellular field distribution in the target area of DBS allows for an investigation of the mechanisms of action of DBS in a simulation model that captures both network and volume conductor effects. Analytic models of the extracellular field distribution add the spatial location and extracellular stimulation of the neurons to the network model, but are unable to consider effects of the geometry of the electrode, the brain anatomy, and the electrode-tissue interface. Therefore, in this study a 3D finite element model of a rodent brain was coupled to a network model to investigate the influence of uncertainty in the conductivity profiles of biological tissue on beta-band suppression in the network model during DBS. The application of the gPC method allowed for a computationally efficient quantification of the uncertainty in the quantities of interest of the volume conductor model and the network model on a common workstation, which would be infeasible if standard stochastic methods such as Monte Carlo simulation were used. The computation time and memory consumption for one realization was approximately 2 hours and \SI{3.5}{Gb} on a $2.4\,\SI{}{Ghz}$ core of a common workstation, with over \SI{95}{\percent} of the computation time utilized by the network model. The computation of the realizations for all tissue conductivity samples and stimulation amplitudes required less than \rev{$10$} days on $12 \times 2.4\,\SI{}{Ghz}$ cores of the workstation. A Monte Carlo simulation with comparable accuracy, which would have required a large number of several thousand random samples, would take an estimated computational time of several months and a substantially larger amount of storage. Further, the surrogate models determined by the gPC method of the investigated quantities of interest consist of a series expansion of multi-variate polynomials, which are evaluable for a large number of random samples by Monte Carlo simulation in a comparably small amount of time.\newline \indent
\rev{The window of stimulation amplitudes, at which the level of beta suppression is sensitive to uncertainty in the electrical properties of brain and encapsulation tissue, ranges from \SI{110}{\micro A} to \SI{185}{\micro A} (Fig. \ref{fig:stim_uq_bars}) for the \rev{assumed} uncertainty in the tissue conductivities of the volume conductor model (Table \ref{tab:tissues}).} The stimulation amplitude required to effectively modify the spiking patterns in the neurons of the network model, and subsequently the level of beta suppression, depends not only on the tissue conductivity, but also on various factors such as the exact location of the neurons in the proximity of the electrode, \rev{the properties of the target neuron axons,} the electrode geometry, and the modeling of the encapsulation layer. \rev{Despite these factors, similar stimulation amplitudes to suppress PD symptoms in experimental studies in rodents have been reported with values between \SI{30} and \SI{100}{\micro A} for a \SI{130}{Hz} DBS stimulus effective \cite{so2012, spieles2010}.} In \cite{spieles2010}, an effective stimulation radius of \SI{250}{\micro m} for a stimulation amplitude of \SI{100}{\micro A} was reported. The \rev{axon} collaterals in the present study were positioned within a similar radius around the stimulation electrode. The magnitude of the beta-band oscillatory activity within the network was substantially reduced at stimulation amplitudes \rev{for which not all STN neurons were directly entrained by the DBS} (Fig. \ref{fig:power_spectra_mean}). This result suggests that the estimation of the ``volume of tissue activated'' alone  to predict neural activation during DBS \cite{butson2005, grant2010, schmidt2013} may also overestimate the required stimulus amplitude to obtain therapeutic effects of DBS\rev{, as the propagation of stimulation effects through the connections within the cortico-basal ganglia network is not considered.}\newline \indent
The volume conductor model of a rodent's brain used in this study consists of a DBS electrode, which is surrounded by an encapsulation tissue layer and is placed within the region of the STN, and the bulk brain tissue medium. The model allows for computing \rev{the field distribution} around the electrode with respect to the influence of the encapsulation and brain tissue. Given that this medium is homogeneous, the extracellular potential around the cylindrical electrode resembles an almost spherical field distribution with increasing distance to the electrode with the exterior boundary of the rodent brain having most likely only a minor influence. Nevertheless, neural elements can be located very close to the encapsulation layer, as it is the case in this study, which requires a sufficiently detailed model of the electrode and the neighboring areas. In addition, the generated 3D rodent brain model allows arbitrary orientations and locations for the electrode and neuronal elements inside the brain to be chosen and the brain structures to be incorporated. In the future, the model will be extended with models of basal ganglia brain structures, which will result in heterogeneous and anisotropic electrical properties of the corresponding tissue types within the model and, therefore, result in a complex, non-spherical field distribution.\newline \indent
\rev{The network model used in this study focuses on the STN-GPe loop and cortical inputs to the STN. The application of DBS to the model is based upon the hypothesis that DBS exerts its effects, at least partially, through the activation of STN afferent fibers projecting from the cortex \cite{gardinaru2009}. In this model, DBS activates corticofugal fibre afferent projecting from the cortex to the STN. The resulting activity antidromically modulates the firing patterns of the cortical neurons while also orthodromically activating the single compartment STN neurons.} Previous computational studies have simulated the effect of extracellular potential on isolated neurons \cite{grill2004, mcintyre2004} or used a simplified point source approximation to apply the potential to the network \cite{kang2014}. \rev{Though the present study is the first to combine a detailed volume conductor model with a network model, it is likely that DBS may also effect other neurons and network circuits, not included here, either directly or indirectly \cite{javor2015}. In particular, the question as to whether DBS results in excitation or inhibition of STN neurons remains a controversial topic with both excitatory and inhibitory effects reported \cite{hammond2008}, alongside the possibility that DBS may inhibit activity of the STN soma while simultaneously activating the afferent output of the STN \cite{mcintyre2004}. There is greater agreement regarding the effects of DBS on the afferent output of the STN, with it generally accepted that STN DBS results in stimulation of the STN output time-locked to the stimulus \cite{deniau2010}. Previous models of the corticobasal ganglia network have stimulated STN DBS through the intracellular stimulation of single compartment STN neurons \cite{kang2013, kumaravelu2016, hahn2010, rubin2004}. The single compartment representation of STN neurons in the present study approximates the ``input-output'' characteristics of the STN neurons and does not rule out the possibility that the cell-body of the STN neurons may be inhibited while the nerve axons are simultaneously stimulated. Multi-compartment models of these nuclei, especially of the STN, would increase the physiological accuracy of the model. However, this would also substantially increase the computational time and the use of computational resources. For example, the model used in this study considered in total $1,950$ neural segments along the cortical neurons for extracellular stimulation and required for one realization approximately \SI{3.5}{Gb} memory. Replacing the STN neurons in this study by multi-compartment models, as used for example in \cite{butson2005}, would require extracellular stimulation of additional $6,630$ neural segments, which would result in a substantial increase of the required computational resources and computational time. The number of neurons in each nucleus of the network model, while consistent with what has been used in previous computational studies \cite{rubin2004, terman2002}, is lower than it is physiologically realistic.} The relatively low numbers of neurons, which are highly interconnected throughout the basal ganglia-network, make the model sensitive to changes in electric potential as even small changes in potential can quickly increase the percentage of neurons activated.\newline \indent
The development of realistic volume conductor models coupled to network models of the basal ganglia move towards the prediction of required stimulation amplitudes to obtain therapeutic effects of DBS, based on realistic brain anatomy and electrical properties as well as the electrode-tissue-interface and neural dynamics. For this problem class and computational models, the results of this study are of crucial importance, since they show the substantial influence of brain tissue uncertainty on pathological neural oscillations which are one possible objective function of the corresponding optimization problem. Therefore, an objective function omitting this uncertainty may not sufficiently render patient-specific stimulation profiles, but only represent one possible solution.\newline \indent
Regarding the minor influence of the uncertainty in encapsulation tissue conductivity on the level of beta suppression and the extracellular potential, the results of this study are in agreement with other computational and experimental studies concerning current-controlled DBS \cite{butson2005, grant2010, lempka2010}. For voltage-controlled stimulation, a more precise knowledge of the encapsulation layer properties in the post-operative state may be crucial for modeling the mechanisms of action of DBS in the rodent's brain \cite{schmidt2013}.

\section{Conclusion}
Uncertainty in the electrical properties of biological tissues, including grey and white matter and the encapsulation tissue which forms around the electrode, represents a significant limitation of existing models of the electric field in the vicinity of the DBS electrode. The potential impact of this uncertainty on the dynamics of neural firing and pathological neural oscillations within the cortico-basal ganglia network during DBS is unclear. \rev{The methodology presented in this paper offers a computationally efficient method with which to quantify these effects by coupling a finite element model of the electric field with a physiologically-based network model, and applying a stochastic approach based on the generalized Polynomial Chaos.} The results of this study indicate that modest variations in the electrical conductivity of brain tissue of $13 - 15$ \SI{}{\percent}, can result in substantial deviations, of up to $300$ \SI{}{\percent} about the mean, in the level of beta suppression among STN neurons during DBS. The methods proposed offer an efficient method with which to incorporate known uncertainties in parameters of interest in models of DBS.  

\bibliographystyle{IEEEtran}
\bibliography{uq_tdcs}

\begin{thebibliography}{10}
\providecommand{\url}[1]{#1}
\csname url@samestyle\endcsname
\providecommand{\newblock}{\relax}
\providecommand{\bibinfo}[2]{#2}
\providecommand{\BIBentrySTDinterwordspacing}{\spaceskip=0pt\relax}
\providecommand{\BIBentryALTinterwordstretchfactor}{4}
\providecommand{\BIBentryALTinterwordspacing}{\spaceskip=\fontdimen2\font plus
\BIBentryALTinterwordstretchfactor\fontdimen3\font minus
  \fontdimen4\font\relax}
\providecommand{\BIBforeignlanguage}[2]{{%
\expandafter\ifx\csname l@#1\endcsname\relax
\typeout{** WARNING: IEEEtran.bst: No hyphenation pattern has been}%
\typeout{** loaded for the language `#1'. Using the pattern for}%
\typeout{** the default language instead.}%
\else
\language=\csname l@#1\endcsname
\fi
#2}}
\providecommand{\BIBdecl}{\relax}
\BIBdecl

\bibitem{litvak2011}
V.~Litvak, A.~Jha, A.~Eusebio, R.~Oostenveld, T.~Foltynie, P.~Limousin,
  L.~Zrinzo, M.~Hariz, K.~Friston, and P.~Brown, ``Resting oscillatory
  cortico-subthalamic connectivity in patients with parkinson's disease,''
  \emph{Brain}, vol. 134, pp. 359--374, Feb 2011.

\bibitem{eusebio2008}
A.~Eusebio, C.~C. Chen, C.~S. Lu, S.~T. Lee, C.~H. Tsai, P.~Limousin, M.~Hariz,
  and P.~Brown, ``Effects of low-frequency stimulation of the subthalamic
  nucleus on movement in parkinson's disease,'' \emph{Exp Neurol}, vol. 209,
  pp. 125--130, Jan 2008.

\bibitem{kuhn2008}
A.~A. K\"uhn, F.~Kempf, C.~Br\"ucke, L.~Gaynor~Doyle, I.~Martinez-Torres,
  A.~Pogosyan, T.~Trottenbrg, A.~Kupsch, G.~H. Schneider, M.~I. Hariz,
  W.~Vandenberghe, B.~Nuttin, and P.~Brown, ``{High-frequency stimulation of
  the subthalamic nucleus suppresses oscillatory beta activity in patients with
  Parkinson's disease in parallel with improvement in motor performance},''
  \emph{J Neurosci}, vol.~28, pp. 6165--6173, Jun 2008.

\bibitem{bronte2009}
H.~Bronte-Stewart, C.~Barberini, M.~M. Koop, B.~C. Hill, J.~M. Henderson, and
  B.~Wingeier, ``{The STN beta-band profile in Parkinson's disease is
  stationary and shows prolonged attenuation after deep brain stimulation},''
  \emph{Exp Neurol}, vol. 215, pp. 20--28, Jan 2009.

\bibitem{rubin2004}
J.~Rubin and D.~Terman, ``High frequency stimulation of the subthalamic nucleus
  eliminates pathological thalamic rhythmicity in a computational model,''
  \emph{J Comput Neurosci}, vol.~16, pp. 211--235, May 2004.

\bibitem{hahn2010}
P.~J. Hahn and C.~C. McIntyre, ``Modeling shifts in the rate and pattern of
  subthalamopallidal network activity during deep brain stimulation,'' \emph{J
  Comput Neurosci}, vol.~28, pp. 425--441, Jun 2010.

\bibitem{kang2013}
G.~Kang and M.~M. Lowery, ``{Interaction of Oscillations, and Their Suppression
  via Deep Brain Stimulation, in a Model of the Cortico-Basal Ganglia
  Network},'' \emph{IEEE T Neur Sys Reh}, vol.~21, pp. 244--253, Mar 2013.

\bibitem{butson2005}
C.~R. Butson and C.~C. McIntyre, ``Tissue and electrode capacitance reduce
  neural activation volumes during deep brain stimulation,'' \emph{Clin
  Neurophysiol}, vol. 116, pp. 2490--2500, Oct 2005.

\bibitem{grant2010}
P.~F. Grant and M.~M. Lowery, ``{Effect of Dispersive Conductivity and
  Permittivity in Volume Conductor Models of Deep Brain Stimulation},''
  \emph{IEEE T Bio-med Eng}, vol.~57, pp. 2386--2393, Oct 2010.

\bibitem{schmidt2013biostec}
C.~Schmidt and U.~van Rienen, ``{Single Frequency Approximation of Volume
  Conductor Models for Deep Brain Stimulation Using Equivalent Circuits},''
  \emph{Proceedings of the International Conference on Bio-inspired Systems and
  Signal Processing}, pp. 38--47, Feb 2014.

\bibitem{gabriel2009}
C.~Gabriel, A.~Peyman, and E.~H. Grant, ``{Electrical conductivity of tissue at
  frequencies below 1 MHz},'' \emph{Phys Med Biol}, vol.~54, pp. 4863--4878,
  Aug 2009.

\bibitem{schmidt2013}
C.~Schmidt, P.~Grant, M.~Lowery, and U.~van Rienen, ``{Influence of
  Uncertainties in the Material Properties of Brain Tissue on the Probabilistic
  Volume of Tissue Activated},'' \emph{IEEE Trans Biomed Eng}, vol.~60, pp.
  1378--1387, May 2013.

\bibitem{kang2014}
G.~Kang and M.~M. Lowery, ``{Effects of antidromic and orthodromic activation
  of STN afferent axons during DBS in Parkinson's disease: a simulation
  study},'' \emph{Front Comput Neurosci}, vol. 57. article 32, Mar 2014.

\bibitem{schmidt2015ner}
C.~Schmidt, E.~Dunn, M.~Lowery, and U.~van Rienen, ``{Simulating the
  Therapeutic Effects of Deep Brain Stimulation in Rodents Using a
  Cortico-Basal Ganglia Network and Volume Conductor Model},''
  \emph{Proceedings of the 7th International IEEE EMBS Conference on Neural
  Engineering (NER)}, pp. 852--855, Jun 2015.

\bibitem{otsuka2004}
T.~Otsuka, T.~Abe, T.~Tsukagawa, and W.~J. Song, ``{Conductance-based model of
  the voltage-dependent generation of a plateau potential in subthalamic
  neurons},'' \emph{J Neurophysiol}, vol.~92, pp. 255--264, Jul 2004.

\bibitem{izhikevich2003}
E.~M. Izhikevich, ``{Simple Model of Spiking Neurons},'' \emph{IEEE T Neural
  Networks}, vol.~14, pp. 1569--1572, 2003.

\bibitem{pospischil2008}
M.~Pospischil, M.~Toledo-Rodriguez, C.~Monier, Z.~Piwkowska, T.~Bal,
  Y.~Fregnac, H.~Markram, and A.~Destexhe, ``{Minimal Hodgkin-Huxley type
  models for different classes of cortical and thalamic neurons},'' \emph{Biol
  Cybern}, vol.~99, pp. 427--441, Nov 2008.

\bibitem{foust2011}
A.~J. Foust, Y.~Yu, M.~Popovic, D.~Zecevic, and D.~A. McCormick, ``{Somatic
  membrane potential and Kv1 channels control spike repolarization in cortical
  axon collaterals and presynaptic boutons},'' \emph{J Neurosci}, vol.~26, pp.
  15\,490--15\,498, Oct 2011.

\bibitem{shepherd2013}
G.~M.~G. Shepherd, ``{Corticostriatal connectivity and its role in disease},''
  \emph{Nat Rev Neurosci}, vol.~14, pp. 1030--1041, Jul 2014.

\bibitem{albin1989}
R.~L. Albin, A.~B. Young, and J.~B. Penney, ``{The functional anatomy of basal
  ganglia disorders},'' \emph{Trends Neurosci}, vol.~12, pp. 366--375, Oct
  1989.

\bibitem{li2012}
Q.~Li, Y.~Ke, D.~C. Chan, Z.~M. Qian, K.~K. Yung, H.~Ko, G.~W. Arbuthnott, and
  W.~H. Yung, ``{Therapeutic deep brain stimulation in Parkinsonian rats
  directly influences motor cortex},'' \emph{Neuron}, vol.~76, pp. 1030--1041,
  Dec 2012.

\bibitem{dunn2013}
E.~M. Dunn and M.~M. Lowery, ``{Simulation of PID control schemes for
  closed-loop deep brain stimulation},'' \emph{6th International IEEE/EMBS
  Conference on Neural Engineering (NER)}, pp. 1182--1185, Nov 2013.

\bibitem{hines2003}
M.~L. Hines and N.~T. Carnevale, ``{The NEURON simulation environment},''
  \emph{Neural Comput}, vol.~9, pp. 1179--1209, Aug 1997.

\bibitem{brown2000}
P.~Brown, ``{Cortical drives to human muscle: the Piper and related rhythms},''
  \emph{Prog Neurobiol}, vol.~60, pp. 97--108, Jan 2000.

\bibitem{ma2005}
Y.~Ma, P.~R. Hof, S.~C. Grant, S.~J. Blackband, R.~Bennet, L.~Slatest, M.~D.
  McGuigan, and H.~Benveniste, ``{A three-dimensional digital atlas database of
  the adult C57BL/6J mouse brian by magnetic resonance microscopy},''
  \emph{Neuroscience}, vol. 135, pp. 1203--1215, Sep 2005.

\bibitem{dehaas2012}
G.~G. de~Haas, ``{Towards the neurobiology of compulsive rituals},'' \emph{PhD
  Thesis. BOXPress Oisterwijk}, p.~81, 2012.

\bibitem{gimsa2005}
J.~Gimsa, B.~Habel, U.~Schreiber, U.~van Rienen, U.~Strauss, and U.~Gimsa,
  ``{Choosing electrodes for deep brain stimulation experiments -
  electrochemical considerations},'' \emph{J Neurosci Meth}, vol. 142, pp.
  251--265, Mar 2005.

\bibitem{yousif2009}
N.~Yousif and X.~Liu, ``Investigating the depth electrode-brain interface in
  deep brain stimulation using finite element models with graded complexity in
  structure and solution,'' \emph{J Neurosci Meth}, vol. 184, no.~1, pp.
  142--151, Oct 2009.

\bibitem{gabriel1996}
S.~Gabriel, R.~W. Lau, and C.~Gabriel, ``{The dielectric properties of
  biological tissues: {III} {P}arametric models for the dielectric spectrum of
  tissues},'' \emph{Phys Med Biol}, vol.~41, pp. 2271--2293, Nov 1996.

\bibitem{gradinaru2009}
V.~Gradinaru, M.~Mogri, K.~R. Thompson, J.~M. Henderson, and K.~Deisseroth,
  ``{Optical deconstruction of parkinsonian neural circuitry},''
  \emph{Science}, vol.~17, pp. 354--359, Apr 2009.

\bibitem{astrom2015}
M.~\r{A}str\"om, E.~Diczfalusy, H.~Martens, and K.~W\r{a}rdell, ``{Relationship
  between Neural Activation and Electric Field Distribution during Deep Brain
  Stimulation},'' \emph{IEEE Trans Bio Med Eng}, vol.~62, pp. 664--672, Feb
  2015.

\bibitem{xiu2010}
D.~Xiu, \emph{{Numerical Methods for Stochastic Computations: {A} Spectral
  Method Approach}}.\hskip 1em plus 0.5em minus 0.4em\relax Princeton
  University Press, 2010.

\bibitem{constantine2012}
P.~G. Constantine, M.~S. Eldred, and E.~T. Phipps, ``Sparse pseudospectral
  approximation method,'' \emph{Comput Method Appl M}, vol. 229–232, pp. 1 --
  12, July 2012.

\bibitem{nobile2008}
F.~Nobile, R.~Tempone, and C.~G. Webster, ``A sparse grid stochastic
  collocation method for partial differential equations with random input
  data,'' \emph{SIAM J Numer Anal}, vol.~46, pp. 2309--2345, May 2008.

\bibitem{sobol2001}
I.~M. Sobol, ``{Global sensitivity indices for nonlinear mathematical models
  and their Monte Carlo estimates},'' \emph{Ma Comput Sci Eng}, vol.~55, pp.
  271--280, Feb 2001.

\bibitem{schmidt2015jne}
C.~Schmidt, S.~Wagner, M.~Burger, U.~van Rienen, and C.~H. Wolters, ``{Impact
  of Uncertain Head Tissue Conductivity in the Optimization of Trancranial
  Direct Current Stimulation for an Auditory Target},'' \emph{J Neural Eng},
  vol. 12. art. no. 046028, Aug 2015.

\bibitem{schmidt2015jbhi}
C.~Schmidt, U.~Zimmermann, and U.~van Rienen, ``{Modeling of an Optimized
  Electro-Stimulative Hip Revision System Under Consideration of Uncertainty in
  the Conductivity of Bone Tissue},'' \emph{IEEE J Biomed Heal Inf}, vol.~19,
  pp. 1321--1330, Jun 2015.

\bibitem{kuhn2009}
A.~A. K\"uhn, A.~Tsui, T.~Aziz, N.~Ray, C.~Br\"ucke, A.~Kupsch, G.~H.
  Schneider, and P.~Brown, ``{Pathological synchronisation in the subthalamic
  nucleus of patients with Parkinson's disease relates to both bradykinesia and
  rigidity},'' \emph{Exp Neurol}, vol. 215, pp. 380--387, Feb 2009.

\bibitem{hammond2007}
C.~Hammond, H.~Bergman, and P.~Brown, ``{Pathological synchronization in
  Parkinson's disease: networks, models and treatments},'' \emph{Trends
  Neurosci}, vol.~30, pp. 357--364, Jul 2007.

\bibitem{so2012}
R.~G. So, G.~C. McConnell, A.~T. August, and W.~M. Grill, ``{Characterizing
  Effects of Subthalamic Nucleus Deep Brain Stimulation on
  Methamphetamine-Induced Circling Behavior in Hemi-Parkinsonian Rats},''
  \emph{IEEE Trans Neural Syst Rehabil Eng}, vol.~20, pp. 626--635, Sep 2012.

\bibitem{spieles2010}
A.~L. Spieles-Engemann, T.~J. Collier, and C.~E. Sortwell, ``{A functionally
  relevant and long-term model of deep brain stimulation of the rat subthalamic
  nucleus: Advantages and considerations},'' \emph{Eur J Neurosci}, vol.~32,
  pp. 1092--1099, Oct 2010.

\bibitem{gardinaru2009}
V.~Gardinaru, M.~Mogri, K.~R. Thompson, J.~M. Henderson, and K.~Deisseroth,
  ``{Optical deconstruction of parkinsonian neural circuitry},''
  \emph{Science}, vol. 324, pp. 354--359, Apr 2009.

\bibitem{grill2004}
W.~M. Grill, A.~N. Snyder, and S.~Miocinovic, ``{Deep brain stimulation creates
  an informational lesion of the stimulated nucleus},'' \emph{Neuroreport},
  vol.~15, pp. 1137--1140, May 2004.

\bibitem{mcintyre2004}
C.~C. McIntyre, W.~M. Grill, D.~L. Sherman, and N.~V. Thakor, ``{Cellular
  effects of deep brain stimulation: model-based analysis of activation and
  inhibition},'' \emph{J Neurophysiol}, vol.~91, pp. 1457--1469, Apr 2004.

\bibitem{javor2015}
B.~N. J\'avor-Duray, M.~Vinck, M.~van~der Roest, A.~B. Mulder, C.~J. Stam,
  H.~W. Berendse, and P.~Voorn, ``{Early-onset cortico-cortical synchronization
  in the hemiparkinsonian rat model},'' \emph{J Neurophysiol}, vol. 113, pp.
  925--936, Feb 2015.

\bibitem{hammond2008}
C.~Hammond, R.~Ammari, B.~Bioulac, and L.~Garcia, ``{Latest view on the
  mechanism of action of deep brain stimulation},'' \emph{Mov Disord}, vol.~23,
  pp. 2111--2121, Nov 2008.

\bibitem{deniau2010}
J.~M. Deniau, B.~Degos, C.~Bosch, and N.~Maurice, ``{Deep brain stimulation
  mechanisms: beyond the concept of local functional inhibition},'' \emph{Eur J
  Neurosci}, vol.~32, pp. 1080--1091, Oct 2010.

\bibitem{kumaravelu2016}
K.~Kumaravelu, D.~T. Brocker, and W.~M. Grill, ``{A biophysical model of the
  cortex-basal ganglia-thalamus network in the 6-OHDA lesioned rat model of
  Parkinson's disease},'' \emph{J Comput Neurosci}, vol.~40, pp. 207--229, Apr
  2016.

\bibitem{terman2002}
D.~Terman, J.~E. Rubin, A.~C. Yew, and J.~Wilson, ``{Activity Patterns in a
  Model for the Subthalamopallidal Network of the Basal Ganglia},'' \emph{J
  Neurosci}, vol.~22, pp. 2963--2976, Apr 2002.

\bibitem{lempka2010}
S.~F. Lempka, M.~D. Johnson, S.~Miocinovic, J.~L. Vitek, and C.~C. McIntyre,
  ``{Current-controlled deep brain stimulation reduces in vivo voltage
  fluctuations observed during voltage-controlled stimulation},'' \emph{Clin
  Neurophysiol}, vol.~12, pp. 2128--2133, Dec 2010.

\end{thebibliography}

\ifCLASSOPTIONcaptionsoff
  \newpage
\fi

\begin{IEEEbiography}[{\includegraphics[width=1in,height=1.25in,clip,keepaspectratio]{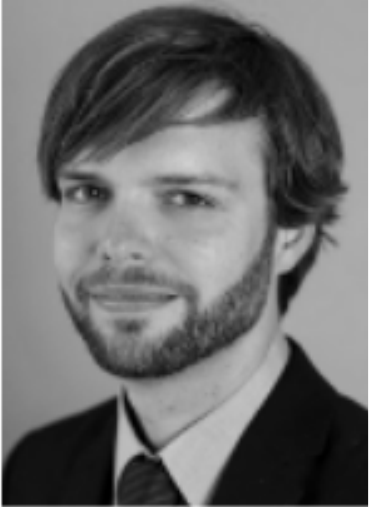}}]
{Christian Schmidt (M'12)}
received his PhD from the Faculty of Computer Science and Electrical Engineering at the University of Rostock in 2013. His field of research comprises computational engineering for several applications ranging from neuroscience and electro-stimulative implants to the design of superconducting radio frequency cavities as used in accelerator facilities. His research interests include also the application and development of stochastical methods for uncertainty quantification.
\end{IEEEbiography}

\begin{IEEEbiography}[{\includegraphics[width=1in,height=1.25in,clip,keepaspectratio]{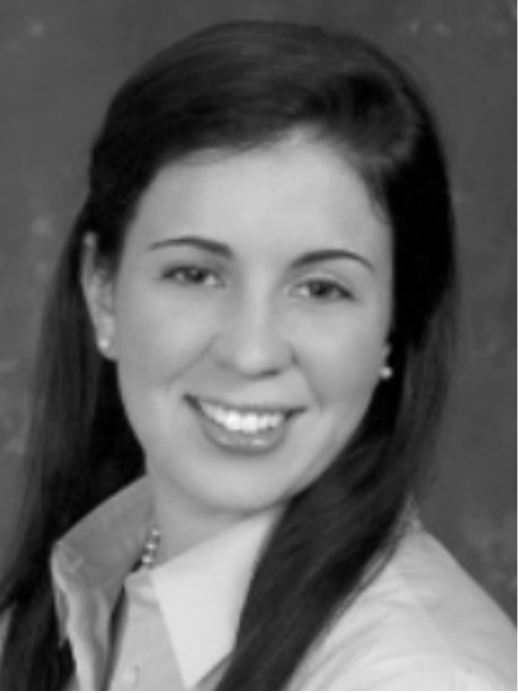}}]
{Eleanor Dunn}
received her B.S. degree in biomedical engineering from Rensselaer Polytechnic Institute, Troy, New York, United States in 2010. She is currently working toward a Ph.D degree in biomedical engineering from University College Dublin, Dublin, Ireland. Her research interests include computational modelling of neural circuits and deep brain stimulation.
\end{IEEEbiography}

\begin{IEEEbiography}[{\includegraphics[width=1in,height=1.25in,clip,keepaspectratio]{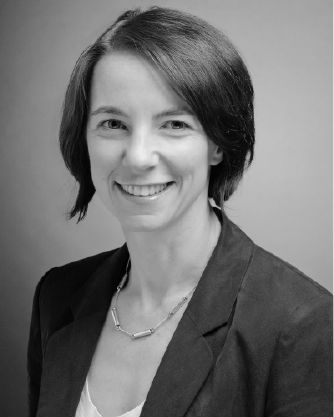}}]
{Madeleine Lowery (M'00)}
is currently an Associate Professor in the School of Electrical and Electronic Engineering, University College Dublin. Her research is focused on using engineering methods to understand the human nervous system as it relates to movement in health and disease, and to design therapies and technologies to improve impaired motor function.  Her research interests include electromyography, myoelectric prosthetic control, bioelectromagnetics, electrical stimulation, deep brain stimulation and neural control of movement. 
\end{IEEEbiography}

\begin{IEEEbiography}[{\includegraphics[width=1in,height=1.25in,clip,keepaspectratio]{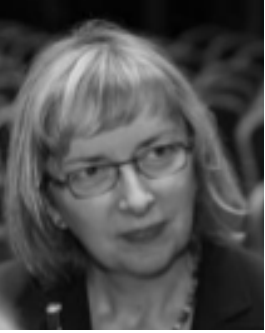}}]
{Ursula van Rienen (M'01)}
received the venia legendi for the fields "Electromagnetic Field Theory" and "Scientific Computing" at the Technische Universit\"at Darmstadt. Since 1997, she holds the chair in "Electromagnetic Field Theory" at the University of Rostock. Her research work is focused on computational electromagnetics with various applications, ranging from biomedical engineering to accelerator physics.
\end{IEEEbiography}




\end{document}